\documentclass[prl,twocolumn,preprintnumbers,superscriptaddress]{revtex4-2}

\usepackage{graphicx}
\usepackage{dcolumn}
\usepackage{bm}
\usepackage{amsmath}
\usepackage{amssymb}
\usepackage{xcolor,dsfont}
\usepackage{times}
\usepackage[colorlinks=true,allcolors=blue]{hyperref}

\newcommand{\para}{\,||\,}
\let\vec\boldsymbol

\begin{document}
\title{Combing the helical phase of chiral magnets with electric currents}
\author{Jan Masell \href{https://orcid.org/0000-0002-9951-4452}{\includegraphics[height=0.75em]{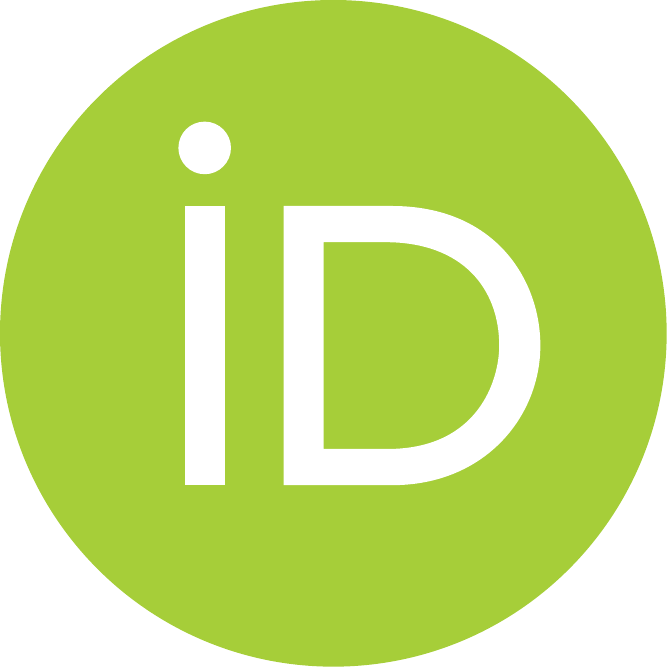}}}
\affiliation{RIKEN Center for Emergent Matter Science (CEMS), Wako, 351-0198, Japan}
\author{Xiuzhen Yu \href{https://orcid.org/0000-0003-3136-7289}{\includegraphics[height=0.75em]{orcid.pdf}}}
\affiliation{RIKEN Center for Emergent Matter Science (CEMS), Wako, 351-0198, Japan}
\author{Naoya Kanazawa \href{https://orcid.org/0000-0003-3270-2915}{\includegraphics[height=0.75em]{orcid.pdf}}}
\affiliation{Department of Applied Physics, University of Tokyo, Tokyo, 113-8656, Japan}
\author{Yoshinori Tokura \href{https://orcid.org/0000-0002-2732-4983}{\includegraphics[height=0.75em]{orcid.pdf}}}
\affiliation{RIKEN Center for Emergent Matter Science (CEMS), Wako, 351-0198, Japan}
\affiliation{Department of Applied Physics, University of Tokyo, Tokyo, 113-8656, Japan}
\affiliation{Tokyo College, University of Tokyo, Tokyo 113-8656, Japan}
\author{Naoto Nagaosa \href{https://orcid.org/0000-0001-7924-6000}{\includegraphics[height=0.75em]{orcid.pdf}}}
\affiliation{RIKEN Center for Emergent Matter Science (CEMS), Wako, 351-0198, Japan}
\affiliation{Department of Applied Physics, University of Tokyo, Tokyo, 113-8656, Japan}

\date{\today}

\begin{abstract}
The competition between the ferromagnetic exchange interaction and anti-symmetric Dzyaloshinskii-Moriya interaction can stabilize a helical phase or support the formation of skyrmions. 
In thin films of chiral magnets, the current density can be large enough to unpin the helical phase and reveal its nontrivial dynamics. 
We theoretically study the dynamics of the helical phase under spin-transfer torques that reveal distinct orientation processes, driven by topological defects in the bulk or induced by edges, limited by instabilities at higher currents.
Our experiments confirm the possibility of on-demand switching the helical orientation by current pulses. 
This helical orientation might serve as a novel order parameter in future spintronics applications.
\end{abstract}

\maketitle

\paragraph*{Introduction --}
In magnetic metals, the magnetization acts on the conduction electrons as a local magnetic field and induces a spin-polarization.
When a current is induced, this coupling has consequences for both the electrons and the magnetization beyond the anomalous Hall effect: 
On the one hand, the spin of the conduction electron locally adapts to the magnetization which can lead to phenomena such as a topological Hall effect from picking up a real-space Berry phase~\cite{Binz2008} or an anisotropic magneto-resistance which reflects the anisotropic magnetic order.
On the other hand, the magnetization can be spatially inhomogeneous and experiences a spin-transfer torque (STT) due to the local reorientation of the spin-polarized current~\cite{Slonczewski1996,Berger1996}.
This electrical control of magnetic states is important for both fundamental research and potential applications.~\cite{Chappert2007} 
For example, it is exploited in commercially available STT-MRAM devices~\cite{Bhatti2017} and can be used to move magnetic domain walls~\cite{Berger1984,Freitas1985} which might lead to shift register memory devices.~\cite{Parkin2008}

More recently, it was found that magnetic skyrmions can be stabilized in chiral magnets~\cite{Bogdanov1989,Muhlbauer2009,Yu2010} and arouse great interest because of   their nanometer size,~\cite{Fert2017,Everschorsitte2018} non-trivial realspace topology,~\cite{Milde2013,Nagaosa2013} and high mobility~\cite{Jonietz2010,Schulz2012,Yu2012,Woo2016} which is interesting for various applications.~\cite{Fert2013,Masell2020b}
In simple chiral ferromagnets like FeGe, spin-orbit coupling induces an antisymmetric Dzyaloshinskii-Moriya exchange interaction (DMI)~\cite{Dzyaloshinskii1958,Moriya1960} which can stabilize skyrmion lattices at certain  magnetic fields and temperatures. 
However, the predominant magnetic phase is not a skyrmion lattice but a topologically trivial (multidomain) helical phase.~\cite{Dzyaloshinskii1965,Bak1980,Uchida2006}
\begin{figure}
\centering{
    \includegraphics[width=\columnwidth]{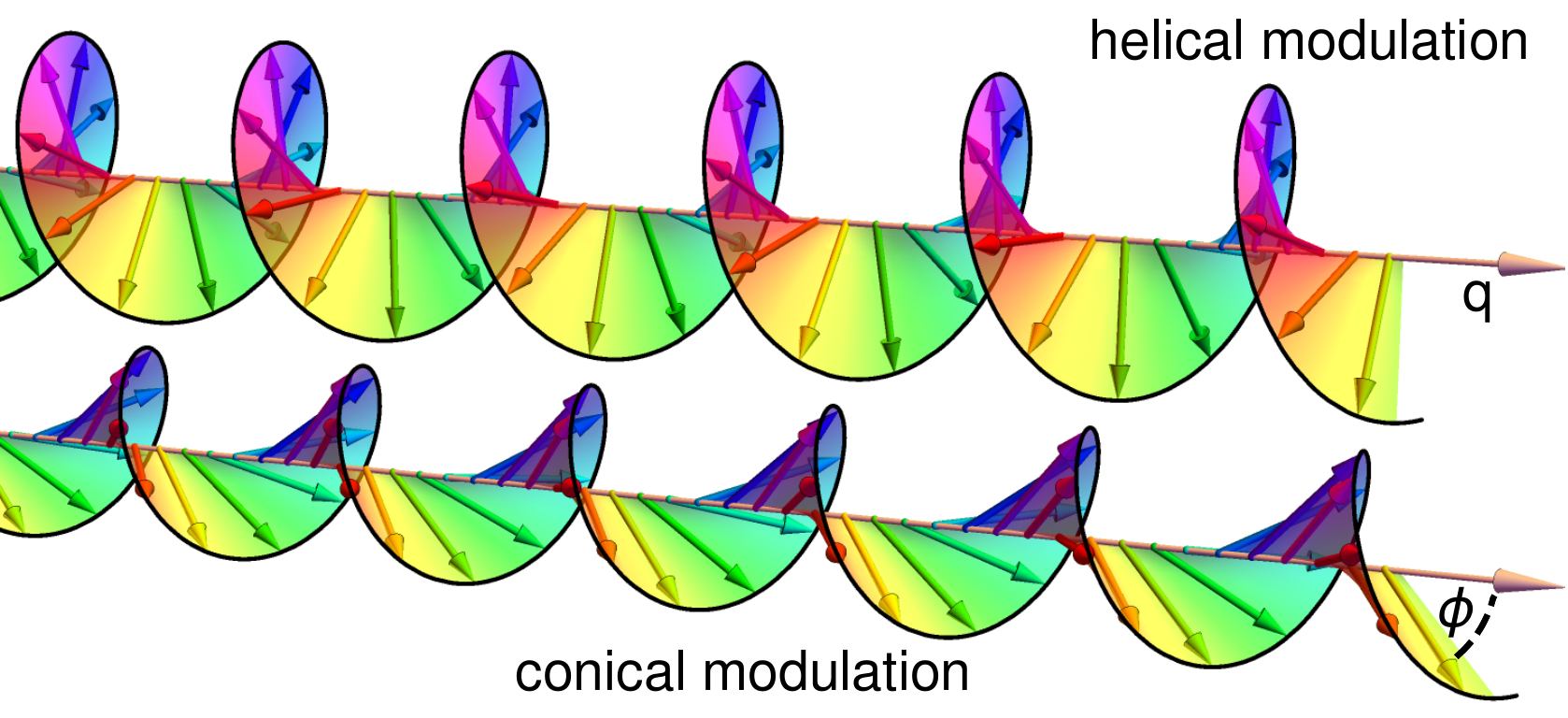}   
    \caption{
    In the helical phase, the magnetization rotates in the plane perpendicular to the $\vec q$-vector.
    Spin-transfer torques drive the helical state out of equilibrium into a conical state where the magnetization tilts towards the $\vec q$-axis.
    }
    \label{fig1}
} 
\end{figure}
Fig.~\ref{fig1} shows how the magnetization in the helical phase winds in the plane perpendicular to the $\vec q$-vector, which defines the orientation of the phase.
When applying a magnetic field $\mathbf{H}$,  $\vec q$ and the orientation of the helical phase can be rotated as $\mathbf{H}\para\vec{q}$ minimizes the energy.~\cite{Bauer2017}
When applying an electric current density $\mathbf{j}$, in turn, the helical phase and its orientation  usually stay pinned.
The reason is that, in contrast to the easily manipulable skyrmion lattice, the helical phase features one extra translation invariant direction perpendicular to its $\vec q$-vector. 
In this direction the helical phase is softer against deformations~\cite{Belitz2006} which leads to stronger pinning at defects~\cite{Iwasaki2013,Hoshino2018} such that very high currents are required for depinning.
\begin{figure*} 
\centering{
    \includegraphics[width=\textwidth]{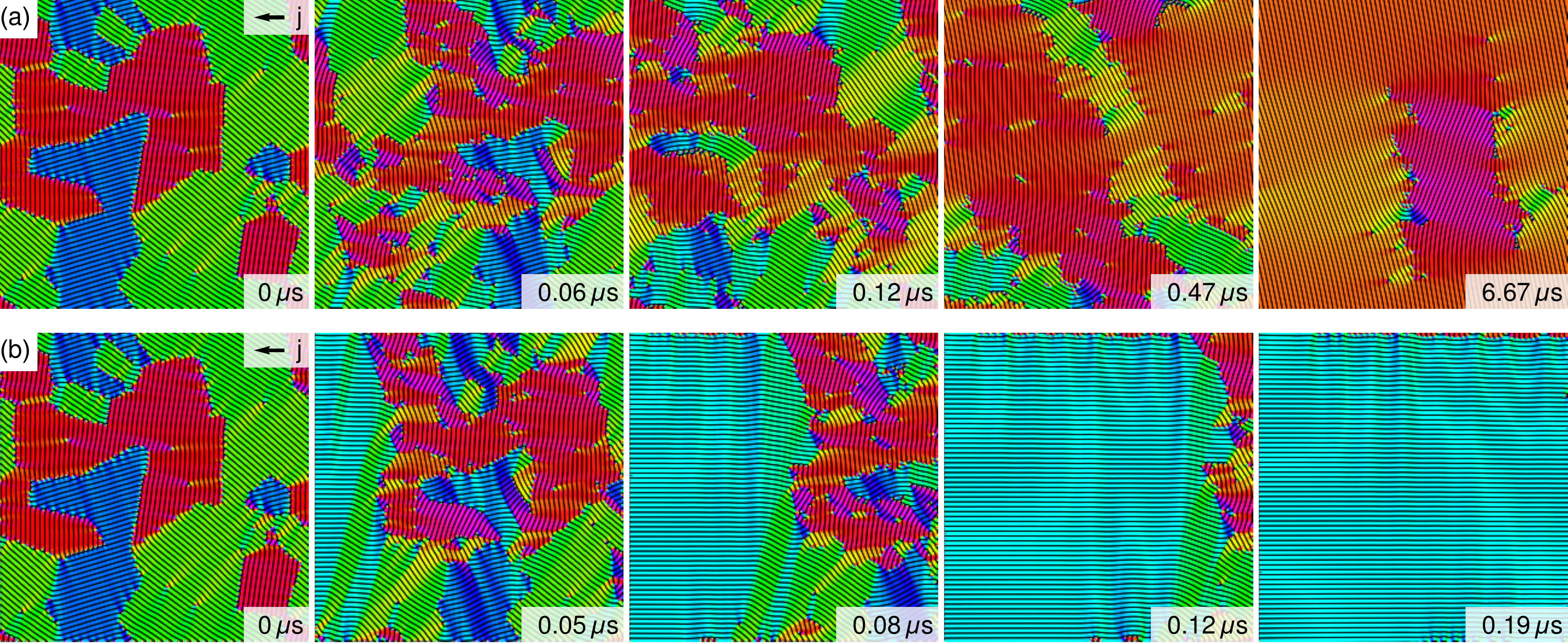} 
    \caption{  
    Snapshots of the magnetization at times $t$ as indicated. 
    (a) With periodic boundary conditions, the dynamics are dominated by defects which order the helix with $\vec q \para \vec j$.
    (b) In a finite size system, the old pattern is pushed out of the system and replaced by a helical phase with $\vec q \perp \vec j$.
    The color encodes the local orientation $\vec{\hat{q}}$ of the helix.
    Additionally, darker color encodes a larger $m_z>0$ and lighter color encodes a larger cone angle.
    Results are obtained for $j=1.6\times10^{11}\mathrm{A}/\mathrm{m}^2$ in a system of size $4.47\times4.47\,\mu\mathrm{m}^2$.  
    }
    \label{fig2} 
}
\end{figure*}
However, even when depinned, the dynamics of the helix are not determined by the state of lowest energy as the system is pumped out of equilibrium.
Instead, its dynamics are governed by the Landau-Lifshitz-Gilbert-Slonczewski equation (LLGS)~\cite{Slonczewski1996,Berger1996,Gilbert2004,Zhang2004}
\begin{equation}
\begin{split}
\frac{d\vec{\hat m}}{dt} =
    &- \gamma\, \vec{\hat m} \times \vec{B}_{\mathrm{eff}} + \alpha\, \vec{\hat m} \times \frac{d\vec{\hat m}}{dt} \\
    &+ \frac{P\mu_\mathrm{B}}{e M_{s}(1+\beta^{2})}\Big(\! \left(\vec{j} \!\cdot\! \nabla\right) \vec{\hat m} - \beta\, \vec{\hat m}\times(\vec{j} \!\cdot\! \nabla ) \vec{\hat m} \Big)\,,
\end{split}
\label{eq:LLGS}
\end{equation}
where $\vec{\hat m} = \vec{M}/M_s$ is the normalized magnetization, $\gamma$ is the gyromagnetic ratio, $\vec{B}_{\mathrm{eff}}$ is the effective magnetic field, $\alpha$ is the Gilbert damping, $P$ is the spin polarization, $e>0$ is the electron charge, $\beta$ is the non-adiabatic damping parameter, and $\vec j$ is the current density.
So far, studies of the dynamics were limited to the time-reversal symmetry breaking effect of the current, which induces a finite cone angle $\phi$,~\cite{Goto2008,Nagaosa2019,Yokouchi2020}
schematically shown in Fig.~\ref{fig1}, irrespective of whether the helix is pinned or mobile.

In this paper, we study the current-induced dynamics of the moving helical phase in chiral magnets.
Our analytical analysis and numerical simulations show a transition from a multidomain to single domain helical phase with $\vec q \para\vec j$ deep in the bulk opposed to $\vec q \perp \vec j$ at the edge of the system.
Various instabilities add to the interesting dynamics.
Our experiments confirm the current-induced reorientation in a thin specimen of FeGe which could be exploited in novel storage devices, e.g., MRAM cells~\cite{Bhatti2017} or memristors~\cite{Yang2013} which measure an orientation dependent resistance.

\paragraph*{Results -- }

The orientation of the helical phase is usually pinned by anisotropies which leads to a multidomain state when cooling below the Curie temperature.~\cite{Bak1980,Bauer2017}
For our theoretical analysis, we consider large current densities such that dynamical effects dominate over such orientational anisotropies or pinning by defects.
We also neglect the effect of the spin-orbit coupling induced torque in chiral magnets.~\cite{Hals2013}
However, the multidomain character turns out to be crucial for the current-induced dynamics.
We therefore model the magnetization far below the Curie temperature by a simple isotropic two-dimensional non-linear sigma model  
\begin{equation}
    E[\vec{\hat m}] = \!\int\!\!\mathrm{d}^2r \,\,
     \left[ \frac{J}{2} (\nabla \vec{\hat m})^2 + D \,\vec{\hat m} \cdot \left(\nabla\times\vec{\hat m}\right) \right]\,,
\end{equation}
where $J=17.5\,\mathrm{pJ/m}$ is the magnetic stiffness and $D=1.58\,\mathrm{mJ/m^2}$ is the DMI in FeGe.~\cite{Beg2015}
As a starting point, we use a multidomain helical phase, see first panels of Fig.~\ref{fig2}, which is prepared via directly minimizing the energy of a tessellation with random orientations of the helix.

\begin{figure}[b]
\centering{
    \includegraphics[width=\columnwidth]{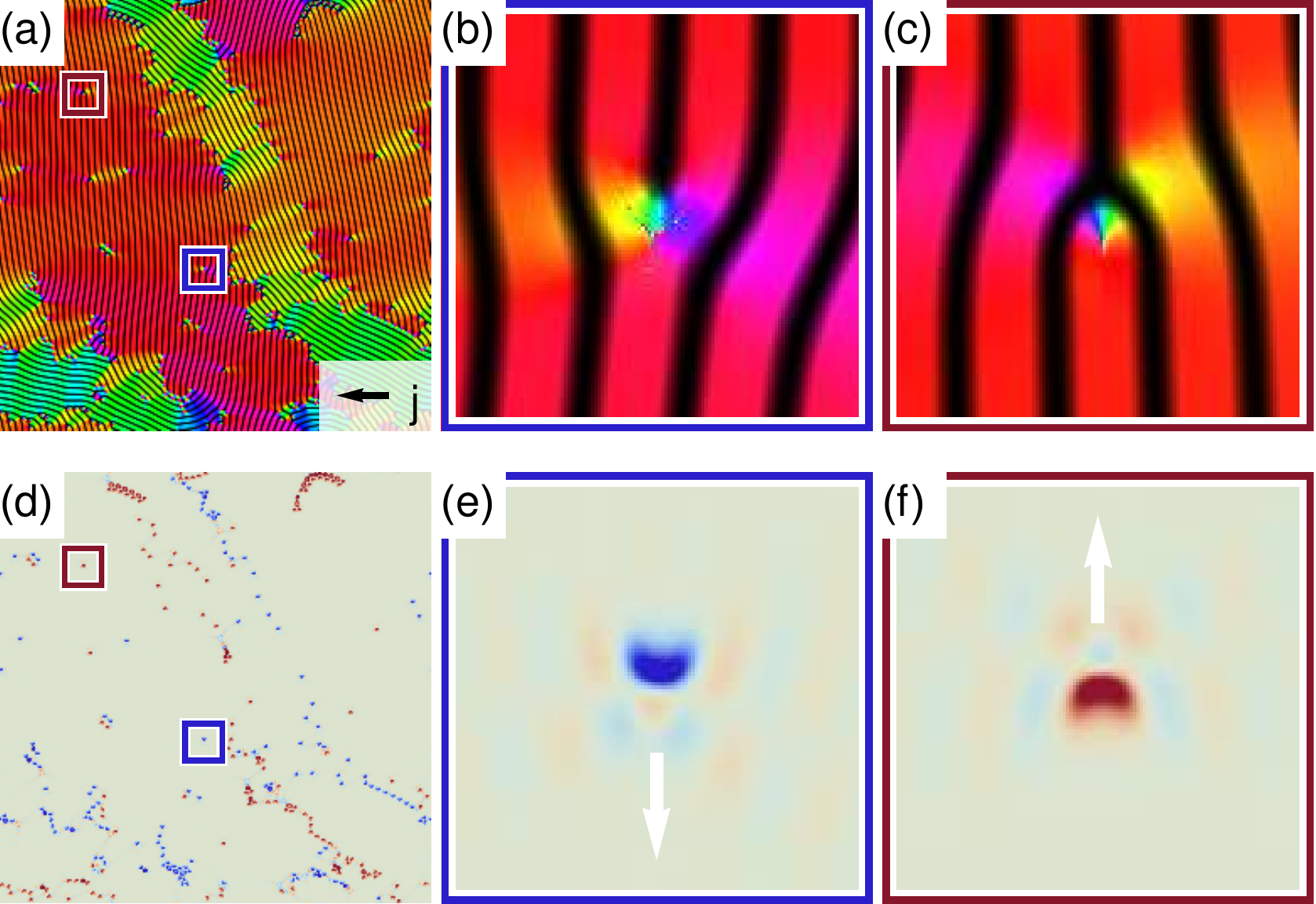}   
    \caption{ 
    The driven helical phase (a) is combed by defects such as the dislocations in (b) (blue frame) and (c) (red frame), using the color code of Fig.~\ref{fig2}.
    Panels (d-f) show the corresponding topological charge density, Eq.~\eqref{eq:2d:w}, with $\mathcal{Q}>0$ (blue)  to $\mathcal{Q}<0$ (red).
    The write arrow in (e) and (f) indicates the Hall motion of the charged defects.
    } 
    \label{fig3}   
} 
\end{figure}

The evolution of the magnetization during our simulations~\cite{footnote1} is shown in Fig.~\ref{fig2}(a) and Movie S1~\cite{Supp} where we apply a current density of $j=1.6\times10^{11}$ $\mathrm{A}/\mathrm{m}^2$ using periodic boundary conditions.
On large timescales, the initially multidomain helical phase transforms into a monodomain phase with $\vec q \para \vec j$.
This ordering process is driven by the dynamics of defects in the helical texture which carry a non-quantized topological charge 
\begin{equation}
 \mathcal{Q} = \int\!\!\!\!\!\int_\Omega \vec{\hat{m}} \cdot \left( \frac{\mathrm{d}\vec{\hat{m}}}{\mathrm{d}x} \times  \frac{\mathrm{d}\vec{\hat{m}}}{\mathrm{d}y} \right) \,\, \mathrm{d}\vec{r}  \in \mathds{R} 
 \label{eq:2d:w}
\end{equation}
and naturally arise at the interfaces between differently oriented helical domains.~\cite{Schoenherr2018} 
\begin{figure*}
\centering{
    \includegraphics[width=\textwidth]{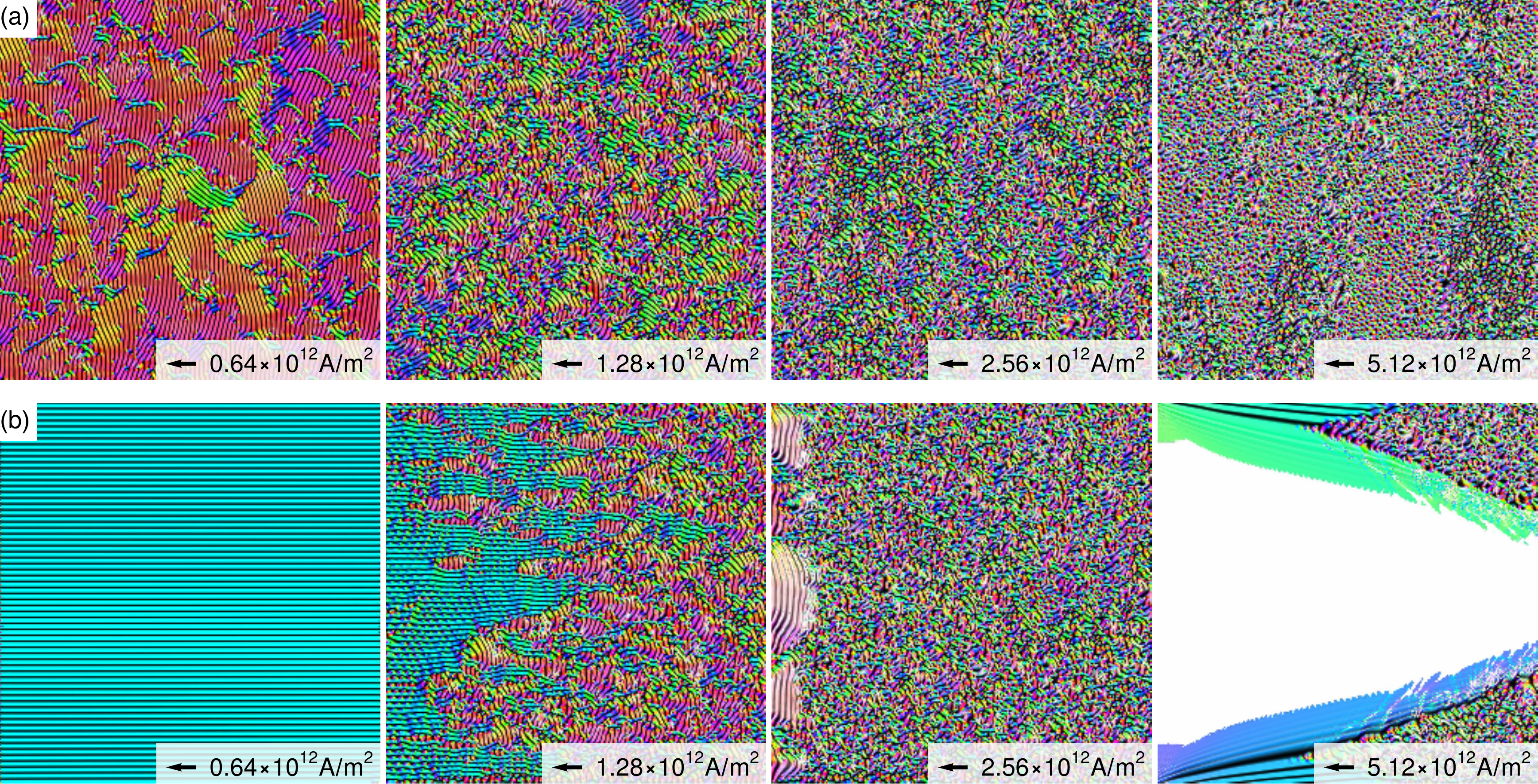}          
    \caption{
    Snapshots of the helical phase after simulating a sufficiently long timespan with (a) periodic boundary condition or (b) open (Neumann) boundary conditions until a steady state is established. 
	The current densities are indicated in each panel.
    Setup and color code are the same as in Fig.~\ref{fig2}.
    In the last panel of (b), the magnetization in the white area is (almost) polarized with $\vec{\hat{m}}=-\vec{\hat{j}}$.
    }  
    \label{fig4} 
}  
\end{figure*}
Here, $\Omega$ is an adequately chosen finite area around the defects which can comprise disclinations~\cite{Schoenherr2018}, dislocations~\cite{Dussaux2016}, and skyrmions~\cite{Muller2017b}.
The topological charge distribution for Fig.~\ref{fig2}(a) (at $t=0.47\,\mu\mathrm{s}$) is shown in Fig.~\ref{fig3}, including a magnified view on a positively and a negatively charged dislocation.
At this relatively small current density, the motion of defects is confined to lanes defined by the helical background.
This background moves uniformly at a velocity $\vec{v} \propto -\vec j$ parallel to the current, whereas a simple Thiele estimate~\cite{Thiele1973} reveils that defects experience a transverse velocity component
\begin{equation}
 \text{sign}(\vec{\hat z} \times \vec v) = - \text{sign}(\alpha-\beta) \, \text{sign}(\mathcal{Q}) \, \text{sign}(\vec j)
\end{equation}
similar to the skyrmion Hall effect.~\cite{Jiang2017,Litzius2017}
This extra transverse velocity is indicated in Fig.~\ref{fig3}(e,f) and can be observed in Movie S2~\cite{Supp}.
Due to their transverse motion, the defects comb their confining lanes such that $\vec q \para \vec j$.
Moreover, oppositely charged defects can annihilate and equally charged defects can form skyrmions that eventually decay under pressure,~\cite{Heil2019} which decreases the number of defects as well as the total winding number $\mathcal{Q}$.
In additional simulations with current densities down to $j=10^{10}$ $\mathrm{A}/\mathrm{m}^2$, the dynamics are slower but qualitatively not different.

At the edges of a finite size system we observe different dynamics:
As shown in Fig.~\ref{fig2}(b) and Movie S3~\cite{Supp}, the collective sliding motion pushes the initial magnetic texture over one edge out of the system.
On the opposite edge, the empty space is filled by a newly entering phase with $\vec q \perp \vec j$ until the entire system is again in a monodomain state.
This is also the case at lower current densities, where the initial texture is only partially expelled from the system. 
In Fig.~\ref{fig2}(b), the current density is large enough to expel almost all of the
initial texture until the entire system is in a monodomain state.
Exceptions are observed at the transverse edges where defects might enter because of their charge-induced dynamics.
We confirmed the order $\vec q \perp \vec j$ also for other orientations of the current relative to the edge.


For larger current densities, the helical phase becomes unstable.
One critical current is set by the analogue of the \emph{Walker breakdown}~\cite{Schryer1974} of magnetic domain walls, which closes the cone angle $\phi\to0$, see Fig.~\ref{fig1}.
The corresponding fixed point of the LLGS equation, Eq.~\eqref{eq:LLGS}, yields the critical current
\begin{equation}
 j_\mathrm{c}^\mathrm{Walker} = \frac{\alpha\,(1+\beta^2)}{|\alpha-\beta|} \frac{\gamma \,e}{\mu_\mathrm{B} } \,(2D-Jq)
\end{equation}
which for FeGe evaluates to the orientation dependent critical current $\vec{j}_\mathrm{c}^\mathrm{Walker} \cdot \vec{\hat{q}} \approx 2.5 \times 10^{12}\,\mathrm{A}/\mathrm{m}^2$ if the wavelength is the equilibrium wavelength $\lambda\approx70\,\mathrm{nm}$.
A stretched wavelength above the equilibrium value increases the critical current up to a factor two whereas decreasing the wavelength reduces the critical current.
Detailed calculations~\cite{elsewhere} reveal that in the helix with $\vec q \perp \vec j$ longitudinal modes soften already below the Walker breakdown which triggers the reduction of the wavenumber. 
An ideal helix would therefore undergo a series of instabilities to larger wavelengths until it finally saturates at the Walker breakdown.
However, in a more realistic setup with defects, these instabilities can be locally activated.
As a result, defects occasionally detach from their helical ties, leaving the system with only a short-range order, see Fig.~\ref{fig4}(a), first panel, and Movie S4~\cite{Supp}.
At higher currents, more defects proliferate and the helical background becomes more transparent, leading to a gradually shorter ranged order, see Fig.~\ref{fig4}(a), which establishes instead of the inplane polarized state.

Another instability of the driven helical phase owes even more directly to its low-dimensional texture, namely the translational invariance and thus softer excitation spectrum perpendicular to $\vec q$.~\cite{Belitz2006}
In fact, \emph{any} finite current perpendicular to $\vec q$ triggers an instability~\cite{elsewhere} which spontaneously breaks the continuous translational symmetry with
\begin{equation}
 k_\perp \propto \vec j \cdot (\vec{\hat z} \times \vec{ \hat q})
\label{eq:2d:kperp}
\end{equation} 
but the time scale for building up the instability scales only as $(j/j_c)^{-4}$ which is very long for small currents.~\cite{elsewhere}
In Fig.~\ref{fig4}(b) and the corresponding Movies S5-S8~\cite{Supp} we show the steady state magnetization obtained from simulations with successively larger currents for the system known from Fig.~\ref{fig2}(b).
In the first panel, the inherent instability is not observed as its timescale is smaller than the time needed to pass once through the system, similar to Fig.~\ref{fig2}(b).
In the second panel, this instability occurs faster and thus can be observed close to the edge which results in the proliferation of dislocations and skyrmions~\cite{Lin2013,Lin2016}.
In the third panel, where $j>j_\mathrm{c}^\mathrm{Walker}$, the length scale of the inherent instability is too small to be observed. 
Instead, patches of the also unstable but more slowly decaying phase with $\vec q \para \vec j$ eventually burst from the edge and decay into the fluctuating background via the Walker breakdown.
For a current $j>2 j_\mathrm{c}^\mathrm{Walker}$, fourth panel, we finally observe a large scale inplane polarized phase, here shown in white, which seeds at the edge but is unstable against both the seemingly laminar and turbulent phases that enter from the transverse edges.
The final state after turning off the current depends on the time-dependence of the current strength.
However, quickly turning off the current can result in a strongly disordered phase.


\begin{figure}
\centering{
    \includegraphics[width=\columnwidth]{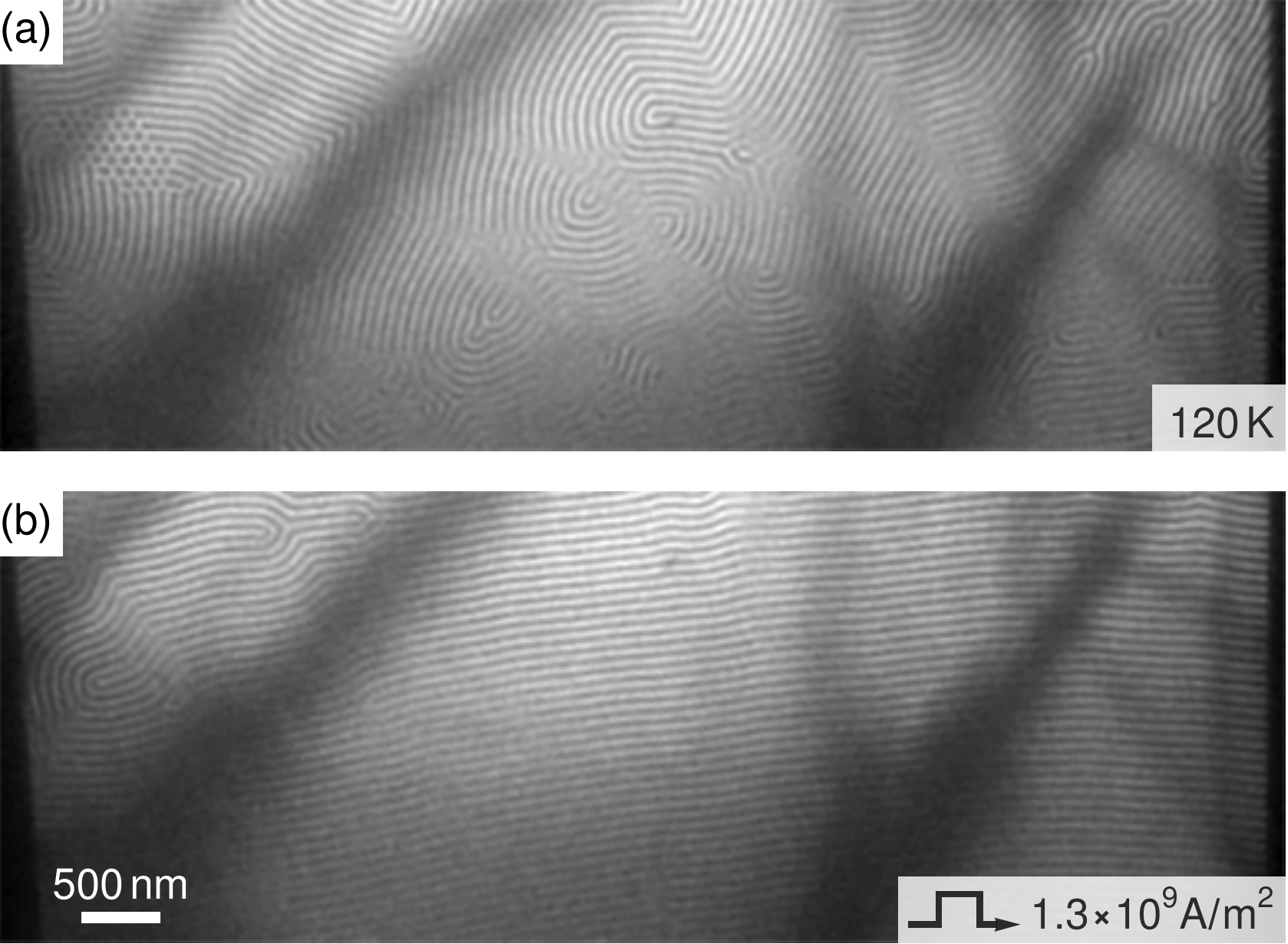} 
    \caption{
    Lorentz TEM images of a $150\,\mathrm{nm}$ thick film of FeGe at $120\,\mathrm{K}$ before  and after a current pulse.
    (a) The initial magnetization shows a multidomain helical phase (stripes) and a small skyrmion cluster (dots on left side).
    (b) After a current pulse with $j = 1.3\times10^9\mathrm{A}/\mathrm{m}^2$ for $0.5\,\mathrm{ms}$ in the horizontal direction the magnetization is in an almost defect-free helical state, ordered with  $\vec{q} \perp \vec{j}$. 
    }
    \label{fig5}
}
\end{figure}

We also experimentally confirm the possibility of current-induced order in the helical phase, using the experimental setup from Ref.~\citenum{Yu2020} where current pulses can be applied through a $20\times20\times0.15\,\mu\mathrm{m}^3$ film of FeGe.
Fig.~\ref{fig5}(a) shows an underfocused Lorentz TEM image of the initial state after cooling to $120\,\mathrm{K}$.
The magnetization appears to be in a multidomain helical phase  and also a skyrmion cluster can be spotted.
After applying a single current pulse of $1.3\times10^9\mathrm{A}/\mathrm{m}^2$ for $0.5\,\mathrm{ms}$, the helical phase is ordered with $\vec q \perp \vec j$ and includes only very few dislocation defects, see Fig.~\ref{fig5}(b).
This observation is in agreement with our theoretical prediction on the edge-induced order for the unpinned helical phase at small currents.
However, we do not observe the defect-induced order $\vec q \para \vec j$ predicted in our simulations as the sample size is much too small.

\paragraph*{Conclusions -- }

In this work, we have analyzed the spin-transfer torque induced dynamics of the helical phase of chiral magnets, how it orders at small currents and how it turns disordered above a critical current.
For large systems, we theoretically predict a reorientation transition from an initially multidomain helical phase to a monodomain phase with $\vec q \para \vec j$, driven by the dynamics of topological defect.
At the one edge of the system, however, we expect a new helical phase with $\vec q \perp \vec j$ to enter.
Our experimental observation in a thin plate of FeGe confirms this edge-induced ordering mechanism.
Above the critical current, where defects are no longer bound to helical lanes, the ordering mechanism in the bulk breaks down but edge-induced order can still be obtained.
However, this edge-induced order is intrincially unstable and shows a cascade of possible instabilities at large currents as shown in Fig.~\ref{fig4}(b).
Albeit our study is focused on chiral magnets, the competition between vertical skyrmion-charge induced motion and parallel translation is expected to be rather ubiquitous.

The current-induced helical orientation can be used to all-electrically imprint an anisotropic pattern onto the magnetization.
Such a pattern shows an anisotropic magnetoresistance dependent on the helical orientation $\vec q$.
This effect might be exploited for novel MRAM-like cells which make use of the helical orientation as an order parameter.
More unconventional ideas can also exploit that the information encoded in the helical orientation is not binary.
In principle, a current can be applied in any direction to realize any orientation of the helix in a device with more than only two logic states.
Moreover, in a larger cell we can use the read-out currents to simultaneously induce fractions of new helical order at the edge while probing the system.
As a result, every read-out operation lowers the resistance of the element which is a key element for memristive computing.~\cite{Yang2013}
Finally, large pulses above the instability can be used to reset the device.
 
In conclusion, the helical phase of chiral magnets seemed featureless compared to magnetic skyrmions which appear in the same class of materials.
We disprove this prejudice, revealing the non-trivial dynamics which will be analyzed further in the future and unleash the helical orientation as a new complex order parameter for future applications.

\begin{acknowledgments}
J.M. thanks V. Kravchuk and M. Garst for the helpful discussions.
This work was financially supported by Grants-In-Aid for Scientific Research (A) (grant no. 18H03676 and 19H00660) from the Japanese Society for the Promotion of Science (JSPS) and the Japan Science and Technology Agency (JST) CREST program (grant number JPMJCR1874).
J.M. acknowledges financial support by JSPS (project No. 19F19815) and the Alexander von Humboldt foundation.
\end{acknowledgments}


\begin{thebibliography}{99.}
\bibitem{elsewhere}
These results will be published elsewhere.
\end{thebibliography}
\end{document}